\documentclass[12pt]{iopart}

\usepackage{graphicx}

\begin{document}

\title{A new numerical method to construct binary neutron star initial data}

\author{Wolfgang Tichy}
\address{Department of Physics, Florida Atlantic University,
             Boca Raton, FL  33431, USA}


%
\newcommand\be{\begin{equation}}
\newcommand\ba{\begin{eqnarray}}

\newcommand\ee{\end{equation}}
\newcommand\ea{\end{eqnarray}}
\newcommand\p{{\partial}}
\newcommand\remove{{{\bf{THIS FIG. OR EQS. COULD BE REMOVED}}}}
%

\begin{abstract}

We present a new numerical method for the generation
of binary neutron star initial data using a method along
the lines of the the Wilson-Mathews or the closely
related conformal thin sandwich approach.
Our method uses six different 
computational domains, which include spatial infinity.
Each domain has its own coordinates which are chosen such that
the star surfaces always coincide with domain boundaries.
These properties facilitate the imposition of boundary conditions.
Since all our fields are smooth inside each domain, we are
able to use an efficient pseudospectral method to solve
the elliptic equations associated with the conformal thin sandwich
approach. Currently we have implemented corotating configurations
with arbitrary mass ratios, but an extension to arbitrary spins
is possible. The main purpose of this paper is to introduce 
our new method and to test our code for several different
configurations.

\end{abstract}



\maketitle

\section{Introduction}

Currently several gravitational wave detectors such as 
LIGO~\cite{LIGO:2007kva,LIGO_web},
Virgo~\cite{VIRGO_FAcernese_etal2008,VIRGO_web}
or GEO~\cite{GEO_web} are already operating,
while several others are in the
planning or construction phase~\cite{Schutz99}. One of the most promising
sources for these detectors are the inspirals and mergers of binary neutron
stars (NS). In order to make predictions about the final phase of such inspirals
and mergers, fully non-linear numerical simulations of the Einstein
Equations are required. To start such simulations initial data are needed.
The emission of gravitational waves
tends to circularize the orbits~\cite{Peters:1963ux,Peters:1964}.
Thus, during the inspiral, we expect the two NSs to be
in quasicircular orbits around each other with a radius
which shrinks on a timescale much larger than the orbital
timescale. This means that the initial data should
have an approximate helical Killing vector $\xi^{\mu}$. 
In addition, one would like to have the initial data in coordinates
such that this approximate symmetry is manifest, i.e. the
time evolution vector should lie along $\xi^{\mu}$, so that the time
derivatives of the evolved quantities are minimized.
In order to achieve these goals we use the 
Wilson-Mathews approach~\cite{Wilson95,Wilson:1996ty}, which is 
closely related to the conformal thin sandwich formalism~\cite{York99}.
The Wilson-Mathews approach approach has already been successfully used by
several groups. Among them are results for  
corotating~\cite{Baumgarte:1997xi,Baumgarte:1997eg,
Mathews:1997pf,Marronetti:1998xv} and 
irrotational~\cite{Bonazzola:1998yq,Gourgoulhon:2000nn,Marronetti:1999ya,
Uryu:1999uu,Marronetti:2003gk} NS binaries with equal masses. 
One group has also produced results for unequal mass 
systems~\cite{Taniguchi:2002ns,Taniguchi:2003hx}.


In this paper we present a new numerical method to construct initial data
for binary NSs in corotating configurations for arbitrary mass ratios.
The main focus is on the numerical method rather than on new physics.
We describe an efficient implementation of this method with the SGRID
code~\cite{Tichy:2006qn}, discuss code tests, and compare with
previous results.

Throughout we will use units where $G=c=1$.
Later when we
present numerical results we will use fully dimensionless
units by setting $\kappa$ in the polytropic equation of
state to $\kappa=G=c=1$.
Latin indices such as $i$ run from 1 to 3,
while Greek indices such as $\mu$ run from 0 to 3.
The paper is organized as follows. 
In Sec.~\ref{BNSequations} we describe the General Relativistic equations
that govern binary neutron stars described by perfect fluids.
Sec.~\ref{num_method} describes our particular numerical implementation
of these equations, followed by results for some particular configurations
in Sec.~\ref{results}. 
We conclude with a discussion of our method in Sec.~\ref{discussion}.

\section{Binary neutron stars in General Relativity}
\label{BNSequations}

In this section we describe the equations governing
binary NSs in quasi circular orbits.

\subsection{ADM decomposition of Einstein's equations}

We use the Arnowitt-Deser-Misner (ADM) decomposition of Einstein's
equations (see e.g.~\cite{Misner73}) and write the line element
\begin{equation}
ds^2 = -\alpha^2 dt^2 + \gamma_{ij} (dx^i + \beta^i)(dx^j + \beta^j) 
\end{equation}
in terms of the lapse $\alpha$, shift $\beta^i$ and the 
3-metric $\gamma_{ij}$. The extrinsic curvature is defined by
\begin{equation}
K_{ij} = -\frac{1}{2\alpha}
         (\partial_t \gamma_{ij} - \pounds_{\beta} \gamma_{ij}),
\end{equation}
With these definitions Einstein's equations split into
the evolution equations
\begin{eqnarray}
\label{evo0}
\partial_t \gamma_{ij} &=& -2\alpha K_{ij} + \pounds_{\beta} \gamma_{ij} \nonumber \\
\partial_t K_{ij} &=& \alpha (R_{ij} - 2 K_{il} K^l_j + K K_{ij})
 - D_i  D_j \alpha + \pounds_{\beta} K_{ij} \nonumber \\
 &&- 8\pi S_{ij} + 4\pi\gamma_{ij}(S-\rho)
\end{eqnarray}
and the Hamiltonian and momentum constraint equations
\begin{eqnarray}
\label{ham0}
R -  K_{ij}  K^{ij} + K^2   &=& 16\pi\rho \nonumber \\
\label{mom0}
D_j(K^{ij} - \gamma^{ij} K) &=& 8\pi j^i .
\end{eqnarray}
Here $R_{ij}$ and $R$ are the Ricci tensor and scalar computed from
$\gamma_{ij}$, $D_i$ is the derivative operator compatible 
with $\gamma_{ij}$ and all indices here are raised and lowered
with the 3-metric $\gamma_{ij}$.
The source terms $\rho$, $j^i$, $S_{ij}$ and 
$S=\gamma^{ij}S_{ij}$ are projections of the stress-energy 
tensor $T_{\mu\nu}$ given by
\begin{eqnarray}
\label{mattervars}
\rho   &=& T_{\mu\nu} n^{\mu} n^{\nu} \nonumber \\
j^i    &=& -T_{\mu\nu} n^{\mu} \gamma^{\nu i} \nonumber \\
S^{ij} &=& T_{\mu\nu} \gamma^{\mu i} \gamma^{\nu j}
\end{eqnarray}
and correspond to the energy density, flux and stress-tensor.
The vector $n^{\mu}$ appearing here is the the 4-vector
normal to a $t=const$ slice.

\subsection{Decomposition of 3-metric and extrinsic curvature}

As in~\cite{Wilson95,Wilson:1996ty}
the 3-metric $\gamma_{ij}$ is decomposed into a
conformal factor $\psi$ and a conformal metric
$\bar{\gamma}_{ij}$ such that
\begin{equation}
\gamma_{ij} = \psi^4 \bar{\gamma}_{ij} .
\end{equation}
The extrinsic curvature is split into its trace $K$ and its
tracefree part $A_{ij}$ by writing it as
\begin{equation}
K_{ij} = A_{ij} + \frac{1}{3} \gamma_{ij} K
\end{equation}

\subsection{Quasi equilibrium assumptions}

We now make some additional simplifying assumptions.
First we assume that our binary is in an approximately circular
orbit and that the stars are corotating.
This implies the existence of an approximate
helical Killing vector $\xi^{\mu}$. In a coordinate system where
this helical symmetry is manifest and the time evolution vector
lies along $\xi^{\mu}$, all time derivatives should approximately be zero.
Here we only assume that the time derivative $\partial_t \bar{\gamma}_{ij}$
of the conformal metric and the time derivative $\partial_t K$ of the trace
of the extrinsic curvature vanish.
The former allows us to express the extrinsic curvature in terms
of the shift and results in
\begin{equation}
A^{ij} = \frac{1}{2\psi^4 \alpha}(\bar{L}\beta)^{ij} , 
\end{equation}
where
\begin{equation}
(\bar{L}\beta)^{ij} = 
\bar{D}^i \beta^j + \bar{D}^j \beta^i -\frac{2}{3}  \bar{D}_k \beta^k ,
\end{equation}
and $\bar{D}_k$ is the derivative operator compatible 
with $\bar{\gamma}_{ij}$.
The assumption $\partial_t K = 0$ together with the evolution equation
of $K$ (derived from Eq.~(\ref{evo0})) implies
\begin{eqnarray}
\label{dK0}
\psi^{-5}[\bar{D}_k\bar{D}^k (\alpha\psi) - \alpha \bar{D}_k\bar{D}^k \psi]
&=& \alpha (R + K)^2 + \beta^i \bar{D}_i K \nonumber \\
& &  +4\pi\alpha(S-3\rho) .
\end{eqnarray}

\subsection{Further simplifications and boundary conditions}

Next we also choose a maximal slice and thus $K=0$, and assume that
the conformal 3-metric is flat and given by~\cite{Wilson95,Wilson:1996ty}
\begin{equation}
\label{conflat}
\bar{\gamma}_{ij} = \delta_{ij}.
\end{equation}
This latter assumption merely simplifies our equations
and could in principle be improved by e.g. choosing a 
post-Newtonian expression for $\bar{\gamma}_{ij}$
as in~\cite{Tichy02,Kelly:2007uc}.
Using Eq.~(\ref{conflat})
the Hamiltonian and momentum constraints in Eq.~(\ref{ham0}) 
and Eq.~(\ref{dK0}) simplify and we obtain
\begin{eqnarray}
\label{ham_mom_dtK0}
\bar{D}^2 \psi &=&
 - \frac{\psi^5}{32\alpha^2} (\bar{L}B)^{ij}(\bar{L}B)_{ij}
 -2\pi \psi^5 \rho \nonumber \\
\bar{D}_j (\bar{L}B)^{ij} &=&
 (\bar{L}B)^{ij} \bar{D}_j \ln(\alpha\psi^{-6})  
 +16\pi\alpha\psi^4 j^i \nonumber \\
\bar{D}^2 (\alpha\psi) &=& \alpha\psi
\left[\frac{7\psi^4}{32\alpha^2}(\bar{L}B)^{ij}(\bar{L}B)_{ij}
      +2\pi\psi^4 (\rho+2S) \right], \nonumber \\
\end{eqnarray}
where
$(\bar{L}B)^{ij} = \bar{D}^i B^j + \bar{D}^j B^i 
- \frac{2}{3} \delta^{ij} \bar{D}_k B^k$,
$\bar{D}_i = \partial_i$, and
\begin{equation}  
B^i = \beta^i + \omega \epsilon^{ij3} (x^j - x_{CM}^j) .
\end{equation}
Here $x_{CM}^i$ denotes the center of mass position and $\omega$
is the orbital angular velocity, which we have chosen to lie along
the $z$-direction. 
The elliptic equations (\ref{ham_mom_dtK0}) have to be solved
subject to the boundary conditions
\begin{equation}
\label{psi_B_alpha_BCs}
\lim_{r\to\infty}\psi = 1, \ \ \ 
\lim_{r\to\infty}B^i = 0, \ \ \
\lim_{r\to\infty}\alpha\psi = 1 .
\end{equation}
at spatial infinity.

\subsection{Matter equations}

We assume that the matter in both stars is a perfect fluid with a
stress-energy tensor
\begin{equation}
T^{\mu\nu} = [\rho_0(1+\epsilon) + P] u^{\mu} u^{\nu} + P g^{\mu\nu}.
\end{equation}
Here $\rho_0$ is the mass density (which is proportional the number
density of baryons), $P$ is the pressure, $\epsilon$ is the internal energy
density divided by $\rho_0$, $u^{\mu}$ is the 4-velocity of the fluid
and $g^{\mu\nu}$ is the spacetime metric.
The matter variables in Eq.(\ref{mattervars}) are then
\begin{eqnarray}
\label{fluid_matter}
\rho   &=& \alpha^2 [\rho_0(1+\epsilon) + P] u^0 u^0 - P \nonumber \\
j^i    &=& \alpha[\rho_0(1+\epsilon) + P] u^0 u^0
           (u^i/u^0 + \beta^i) \nonumber \\
S^{ij} &=& [\rho_0(1+\epsilon) + P]u^0 u^0 
           (u^i/u^0 + \beta^i) (u^j/u^0 + \beta^j) \nonumber \\
       & &  + P \gamma^{ij}
\end{eqnarray}

The fact that $\nabla_{\nu} T^{\mu\nu} =0 $ yields the relativistic
Euler equation
\begin{equation}
[\rho_0(1+\epsilon) + P] u^{\nu} \nabla_{\nu} u^{\mu} 
= -(g^{\mu\nu} + u^{\mu} u^{\nu}) \nabla_{\nu} P, 
\end{equation}
which together with the continuity equation
\begin{equation}
\label{continuity}
\nabla_{\nu} (\rho_0 u^{\nu}) = 0
\end{equation}
governs the fluid.

For corotating stars we can show that the continuity
equation~(\ref{continuity}) is identically satisfied.
Furthermore one can show that the Euler equation leads
to (see e.g. problem 16.17 in~\cite{Lightman75})
\begin{equation}
[\rho_0(1+\epsilon) + P] d \ln(u_{\mu}\xi^{\mu}) = -dP ,
\end{equation}
where $\xi^{\mu}$ is the assumed helical Killing vector.
With the help of the first law of thermodynamics
($d[\rho_0(1+\epsilon)] = [\rho_0(1+\epsilon) + P] d\rho_0 / \rho_0$)
this equation can be integrated to yield
\begin{equation}
\label{Bernoulli}
u_{\mu}\xi^{\mu} = \frac{C_{1/2}\rho_0}{\rho_0(1+\epsilon) + P} ,
\end{equation}
where $C_{1/2}$ are constants of integration for each star. We will
later choose them such that the rest mass of each star has a prescribed value.
In corotating coordinates and taking into account our conformally
flat 3-metric, $u_{\mu}\xi^{\mu}$ can be written as
\begin{equation}
\label{uxi}
u_{\mu}\xi^{\mu} = -1/u^0
= -[\alpha^2 - \psi^4\delta_{ij}\beta^i\beta^j]^{1/2} .
\end{equation}

In order simplify the problem we assume a polytropic equation of state 
\begin{equation}
\label{polytrop}
P = \kappa \rho_0^{1+1/n} .
\end{equation}
It is then convenient to introduce the dimensionless ratio
\begin{equation}
q = P/\rho_0 ,
\end{equation}
which we use to write
\begin{eqnarray}
\label{rhoPS_q}
\rho_0   &=& \kappa^{-n} q^n \nonumber \\
P        &=& \kappa^{-n} q^{n+1} \nonumber \\
\epsilon &=& n q .
\end{eqnarray}

\section{Numerical method}
\label{num_method}


In order to construct binary NS initial data
we have to solve the five elliptic equations in Eq.~(\ref{ham_mom_dtK0}),
with the matter terms given by Eqs.~(\ref{fluid_matter}), (\ref{uxi})
and (\ref{rhoPS_q}). In addition, our data also have to 
satisfy Eq.~(\ref{Bernoulli}), which can be expressed as
\begin{equation}
\label{newq}
q = \frac{1}{n+1}\left(\frac{C_{1/2}}{u_{\mu}\xi^{\mu}} - 1\right) 
\end{equation}
for each star.
We will solve the whole set of equations by iterating over the following
steps:
(i) We first come up with an initial guess for $q$ in 
each star, in practice we simply choose 
Tolman-Oppenheimer-Volkoff (TOV) solutions 
(see e.g. Chap.~23 in~\cite{Misner73}) for each. 
(ii) Next we solve
the 5 coupled elliptic equations~(\ref{ham_mom_dtK0}) for this given $q$.
(iii) Then we use Eq.~(\ref{newq}) to update $q$ in each star.
The constants $C_{1/2}$ in general have different values for each star.
We adjust the value for each star such that it has a prescribed rest
mass. 
After updating $q$ we go back to step (ii) and iterate until
all equations are satisfied up to a given tolerance.

\subsection{Coordinates adapted to star surfaces}

Note that the matter is smooth inside the stars. However,
at the surface (at $q=0$), $\rho_0$, $P$ and $\epsilon$
are not differentiable.
This means that if we want to take advantage of a spectral
method, the star surfaces should be domain boundaries.
A difficulty with our iterative approach, however, is that each time
we update $q$ the matter distributions change, so that 
the stars change shape or even move. Hence the domain boundaries have
to be changed as well.
In order to address this problem we introduce several domains
each with its own coordinates. These coordinates depend on two
freely specifiable functions which will allow us to vary
the location of the domain boundaries, so that we can always
adapt our domains to the current star surfaces in each iteration.
As in the initial data approaches in~\cite{Ansorg:2004ds,Ansorg:2005bp}
our aim was to introduce as few domains as possible.

The coordinates we will use, are very similar to the ones
introduced by Ansorg~\cite{Ansorg:2006gd}. We place both stars on the 
$x$-axis and write down the necessary coordinate transformations in
two steps. First we express the standard Cartesian coordinates as
\begin{eqnarray}
x &=& \frac{b}{2}\left[ \frac{1}{(X^2 + R^2)^2} + 1\right](X^2-R^2)  \nonumber \\
y &=& b\left[ \frac{1}{(X^2 + R^2)^2} - 1\right] X R \cos\phi \nonumber \\
z &=& b\left[ \frac{1}{(X^2 + R^2)^2} - 1\right] X R \sin\phi ,
\end{eqnarray}
where $b$ is a parameter related to the distance between the stars,
and $X$, $R$ are functions of the new coordinates $(A,B,\phi)$
we will use in each domain.
Note that spatial infinity is located at the point where $X=R=0$
and that in order to cover all $(x,y,z)$ it is sufficient
the restrict $X,R,\phi$ to the ranges $0\leq X \leq 1$,
$0\leq R \leq \sqrt{1-X^2}$ and $0\leq \phi \leq 2\pi$.
In order to complete the coordinate transformation between
the Cartesian $(x,y,z)$ and the new coordinates $(A,B,\phi)$,
we now write down $X$ and $R$ as functions of $(A,B,\phi)$.
Inside star1 we use
\begin{eqnarray}
X &=& (1-A)\{\Re[C_{+}(B,\phi)] - B\Re[C_{+}(1,\phi)]\}  \nonumber \\
  & &  + B\cos([1-A]\arg[C_{+}(1,\phi)]) + (1-B)A         \nonumber \\
R &=& (1-A)\{\Im[C_{+}(B,\phi)] - B\Im[C_{+}(1,\phi)]\}   \nonumber \\
  & &  + B\sin([1-A]\arg[C_{+}(1,\phi)]) ,
\end{eqnarray}
where the strictly positive function $\sigma_{+}(B,\phi)$ in
\begin{equation}
C_{+}(B,\phi) = \sqrt{\tanh\left(\frac{\sigma_{+}(B,\phi) +i\pi B}{4}\right)}
\end{equation}
determines the shape of the star surface.
The surface is always located at $A=0$ but depending on the choice for
$\sigma_{+}(B,\phi)$ it will be at different $(x,y,z)$, e.g.
for $\sigma_{+}(B,\phi)=const$ we will get a spherical surface in $(x,y,z)$.
Note that this star is located around $x=b$.
Inside star2 we use a similar transformation given by
\begin{eqnarray}
X &=& (1-A)\{\Re[C_{-}(B,\phi)] - B\Re[C_{-}(1,\phi)]\}   \nonumber \\
  & &  + B\cos(\frac{\pi}{2}A + [1-A]\arg[C_{-}(1,\phi)]) \nonumber \\
R &=& (1-A)\{\Im[C_{-}(B,\phi)] - B\Im[C_{-}(1,\phi)]\}   \nonumber \\
  & &  + B\sin(\frac{\pi}{2}A + [1-A]\arg[C_{-}(1,\phi)]) + (1-B) A , \nonumber \\
\end{eqnarray}
where the strictly negative function $\sigma_{-}(B,\phi)$ in
\begin{equation}
C_{-}(B,\phi) = \sqrt{\tanh\left(\frac{\sigma_{-}(B,\phi) +i\pi B}{4}\right)}
\end{equation}
determines where the star surface ($A=0$) is located in $(x,y,z)$
coordinates. Star2 is located around $x=-b$.
Note that the $A,B,\phi$ coordinates are different inside each star,
but in order to cover each star their ranges are
$0\leq A \leq 1$, $0\leq B \leq 1$ and $0\leq \phi \leq 2\pi$
in each case.

The outside of both stars is covered by two additional domains. 
The first one covers the region outside star1 for all positive $x$,
while the second one covers the region outside star2 for all negative $x$.
Both coordinate transformations can be written as
\begin{eqnarray}
X &=& (1-A)\{\Re[C_{\pm}(B,\phi)] - B\Re[C_{\pm}(1,\phi)]\}   \nonumber \\
  & &  + B\cos(\frac{\pi}{4}A + [1-A]\arg[C_{\pm}(1,\phi)]) \nonumber \\
R &=& (1-A)\{\Im[C_{\pm}(B,\phi)] - B\Im[C_{\pm}(1,\phi)]\}   \nonumber \\
  & &  + B\sin(\frac{\pi}{4}A + [1-A]\arg[C_{\pm}(1,\phi)]) ,
\end{eqnarray}
where we use $C_{+}(B,\phi)$ in the former and $C_{-}(B,\phi)$ in
the latter. In each case the star surface is at $A=0$ and
spatial infinity is at $(A,B)=(1,0)$.

\begin{figure}
\includegraphics[scale=0.91,clip=true]{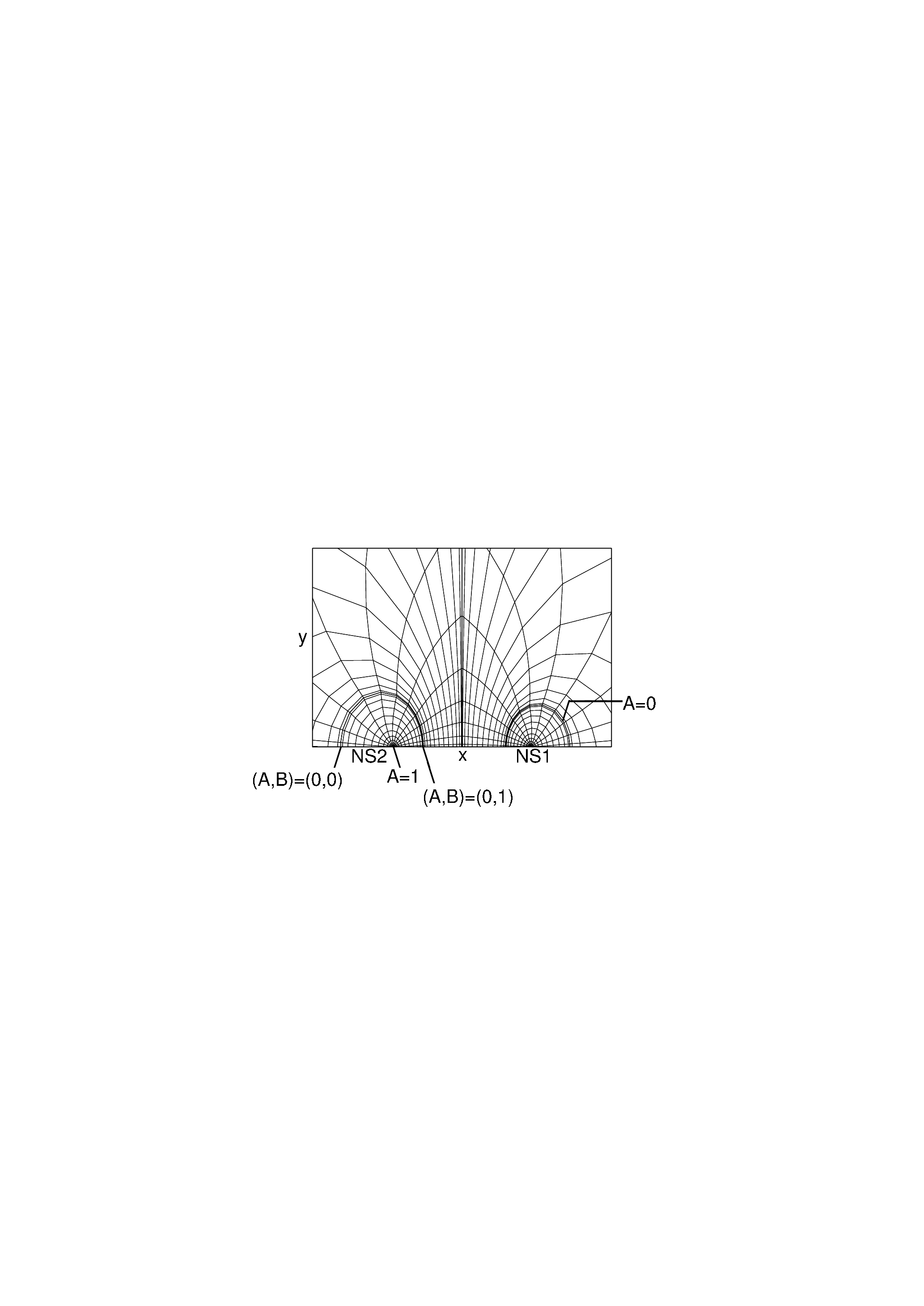}
\caption{\label{coordlines}
The plot shows the lines of constant $A$ and $B$
in the $xy$-plane for $b=1.84$, $\sigma_{+}(B,\phi)=1.51$ and
$\sigma_{-}(B,\phi)=-1.28$. The two neutron stars are marked with 
NS1 and NS2. In addition, the $(x,y)$ positions of a few points
are indicated. The coordinate singularities at $A=1$
inside the stars are located at $x=\pm b$.
}
\end{figure}
Figure \ref{coordlines} shows the coordinate lines in $z=0$ plane.

\subsection{Spectral method}

In order to solve the elliptic equations~(\ref{ham_mom_dtK0})
we use the SGRID code~\cite{Tichy:2006qn} which employs 
pseudospectral methods to accurately compute spatial derivatives.
We use Chebyshev expansions in the $A$- and $B$-directions
and Fourier expansions $\phi$-direction.
As collocation points we choose
\begin{eqnarray}
A_l &=&\frac{1}{2}
       \left[1-\cos\left(\frac{\pi l}{n_A - 1}\right)\right] \nonumber \\
B_j &=&\frac{1}{2}
       \left[1-\cos\left(\frac{\pi j}{n_B - 1}\right)\right] \nonumber \\
\phi_k &=& \frac{2\pi k}{n_{\phi}} , 
\end{eqnarray}
where $l$, $j$, $k$ are integers obeying
\begin{equation}
0 \leq l < n_A, \ \ \
0 \leq j < n_B, \ \ \
0 \leq k < n_{\phi} .
\end{equation}
The number $n_A$, $n_A$ and $n_{\phi}$ of collocation points in each
direction is chosen to be equal in all four domains,
to ensure that the grid points on the boundaries of two adjacent
domains will be at the same $(x,y,z)$ location.
As in~\cite{Tichy:2006qn} will solve Eq.~(\ref{ham_mom_dtK0})
as written down in Cartesian form and compute derivatives
like $\partial_x \psi$ using the chain rule:
\begin{equation}
\label{cartderivs}
\partial_x \psi =
 \frac{\partial A}{\partial x} \partial_A \psi
+\frac{\partial B}{\partial x} \partial_{B} \psi
+\frac{\partial \phi}{\partial x} \partial_{\phi} \psi .
\end{equation}

\begin{figure}
\includegraphics[scale=0.7,clip=true]{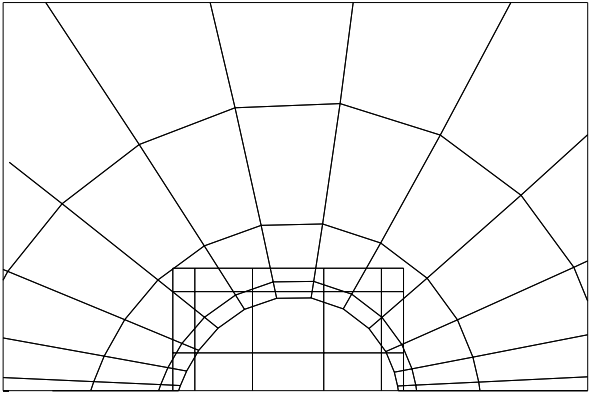}
\caption{\label{cubeinNS1}
In order to avoid the coordinate singularity at 
$A=1$ inside the stars, we have restricted the range of $A$ to
$0 \leq A \leq 0.85$. The remaining space is filled with a small
cube of length $0.016 b$. This figure is a blowup of the inside
of NS1 in Fig. \ref{coordlines} where the small cubes are not visible.
}
\end{figure}

Note that all points with with $B=0$ or $B=1$ lie along
the $x$-axis, with $x$ independent of $\phi$. Hence along
this axis we have the standard coordinate singularity 
of polar coordinates. Furthermore, all points with $A=1$
in the interior of each star 
correspond to just one point on the $x$-axis. Thus there is an
additional coordinate singularity at $A=1$ inside each star.
We have found that if we simply use the $A,B,\phi$ coordinates as described
above we were not able to solve the elliptic equations (\ref{ham_mom_dtK0}).
The culprit is the singularity at $A=1$ inside each star.
Near these points the Jacobian matrix 
$\frac{\partial(A,B,\phi)}{\partial(x,y,z)}$ blows up so strongly 
that we cannot accurately compute derivatives using Eq.~(\ref{cartderivs}).
This problem would also occur if we did use the exact same
coordinates as proposed by Ansorg~\cite{Ansorg:2006gd}. 
In~\cite{Ansorg:2006gd} the problem is not addressed since for black
holes one can use excision boundary conditions and does not need
the inner domains.
One way around this problem would be to construct different basis
functions, which would have to be chosen such that
they have vanishing $B$ and $\phi$ derivatives at $A=1$.
However, then we would not be able to use Fast Fourier transforms anymore
to compute derivatives. For this reason we have chosen a different
approach. We simply restrict the range of $A$ inside each star
so that inside $0 \leq A \leq A_{max}<1$. 
The collocation points in $A$ inside the stars are then given by
\begin{equation}
A_l =\frac{A_{max}}{2}\left[1-\cos\left(\frac{l \pi}{n_{A} - 1}\right)\right].
\end{equation}
We typically choose $A_{max}=0.85$. In this way we completely
avoid the singularities at $A=1$. Of course then, our inner domains
do not cover the entire star interiors any more. Instead they leave out 
a small hole around $A=1$. We simply cover this hole by placing
two additional cubical domains inside each star.
Each cube is chosen such that completely covers the 
hole (described by $A_{max} < A \leq 1$). In each cube we
use standard Cartesian coordinates so that it overlaps with part of the inner
domain covered by the $A,B,\phi$ coordinates.
The collocation points inside each cube are then
\begin{eqnarray}
x_l &=& \frac{x_{min}-x_{max}}{2}\cos\left(\frac{l \pi}{n_{c} - 1}\right)
            +\frac{x_{min}+x_{max}}{2} \nonumber \\
y_j &=& \frac{y_{min}-y_{max}}{2}\cos\left(\frac{j \pi}{n_{c} - 1}\right)
            +\frac{y_{min}+y_{max}}{2} \nonumber \\
z_k &=& \frac{z_{min}-z_{max}}{2}\cos\left(\frac{k \pi}{n_{c} - 1}\right)
            +\frac{z_{min}+z_{max}}{2}, \nonumber \\
\end{eqnarray}
where the minima and maxima $x_{min}$, $x_{max}$, $y_{min}$, $y_{max}$
$z_{min}$, $z_{max}$ are chosen such that the cube covers the hole,
and where $n_c=6$ is typically sufficient, since the cubes are
very small. Figure \ref{cubeinNS1} shows how a cube is fitted into
star1.

In order to numerically solve the elliptic
equations in Eq.~(\ref{ham_mom_dtK0}) we arrange the values
of the fields $\psi$, $B^i$ and $\alpha\psi$ at each grid point
in a vector $w$. Since we solve for 5 fields, the dimension of
this vector is five times the total number of grid points. 
In order to find a $w$ that satisfies Eq.~(\ref{ham_mom_dtK0})
we impose Eq.~(\ref{ham_mom_dtK0}) at all interior grid points.
At adjacent domain boundaries we impose the conditions that fields and
their normal derivatives are equal on both sides. At infinity
we impose Eq.~(\ref{psi_B_alpha_BCs}).
In the domains covered by the $A,B,\phi$ coordinates we impose 
the following regularity conditions along the $x$-axis:
For $k>0$ we demand
\begin{equation}
\Psi(A_l,B_j,\phi_k) = \Psi(A_l,B_j,\phi_0) ,
\end{equation}
while for $k=0$ we impose
\begin{equation}
\partial_{s}\Psi(A_l,B_j,\phi_0) 
+ \partial_{s}\partial_{\phi}\partial_{\phi} \Psi(A_l,B_j,\phi_0) = 0 .
\end{equation}
Here $\Psi$ stands for either $\psi$, $B^i$ or $\alpha\psi$,
and $s=\sqrt{y^2 + z^2}$ is the distance from the $x$-axis.
In order to deal with the cubes which overlap the $A,B,\phi$
covered domains inside each star,
we impose the condition that the fields at points on the cube boundaries
must be equal to the fields in the $A,B,\phi$ covered domain
interpolated to these points. Correspondingly we also
demand that the fields at points on the $A=A_{max}$ boundaries
must equal to the fields in the cube interpolated to these points.
This interpolation is done with our given spectral accuracy.
To compute a field at a point that is not a collocation point,
we first compute the spectral expansion coefficients
from the field values at the collocation points
in the domain of interest. To interpolate to any point
in this domain we then compute the field value from
a sum over coefficients times the basis functions evaluated
at the point in question. This last interpolation step can
be computationally expensive if we interpolate onto many points,
while going between field values and
collocation points can be done via Fast Fourier transforms
and is thus not very expensive. Note, however, that in our
case these interpolations are not too costly, because our
cubical domains have only $6^3$ grid points. 
Such a small number of points should always be sufficient
because the cubic domains are so small so that all fields
are nearly constant inside the cubes, e.g. on the scale
of Fig. 1 the cubes are not visible. This is an advantage
over the domain decomposition used 
in~\cite{Bonazzola:1998qx,Gourgoulhon:2000nn} where interpolations
are needed between domains with many more grid points.

If we take all these conditions into account we obtain 
$N = 4 n_A n_B n_{\phi} + 2n_c^3$ 
non-linear equations of the form
\begin{equation}
f_m (w) = 0, \ \  m=1,2,...,N
\end{equation}
for the $N$ unknowns comprising the solution
vector $w$.
We solve this system of equations by a Newton-Raphson scheme.
This scheme requires an initial guess, for which we simply
use two TOV solutions in conformally flat isotropic coordinates.
In order to solve the linearized equations
\begin{equation}
\label{linearEqs}
\frac{\partial f_m(w)}{\partial w^n } x^n = - f_m (w)
\end{equation}
in each Newton-Raphson step, we note that $f_m (w)$ contains
spectral derivatives of $w$ in different directions, so that
the $N\times N$ matrix $\frac{\partial f_m(w)}{\partial w^n }$ is sparse
in the sense that it contains about 95\% zeros.
So in order to numerically solve the linearized Eq.~(\ref{linearEqs})
we use the sparse matrix solver 
UMFPACK~\cite{Davis-Duff-1997-UMFPACK,Davis-Duff_UMFPACK_1999,
Davis_UMFPACK_V4.3_2004,Davis_UMFPACK_2004,umfpack_web}.

\subsection{Iteration scheme}

As already mentioned solving the elliptic
equations in Eq.~(\ref{ham_mom_dtK0}) once is not enough.
After each solve, we have to adjust $q$ using Eq.~(\ref{newq}).
This adjustment presents the problem that the star
surfaces (located at $q=0$) change. Hence we have to also adjust
$\sigma_{+}(B,\phi)$ and $\sigma_{-}(B,\phi)$ in order
to keep both star surfaces at $A=0$. While this adjustment
is not hard to implement, it incurs a high computational cost
since adjusting $\sigma_{+}(B,\phi)$ and $\sigma_{-}(B,\phi)$ 
amounts to changing our computational grid, and after
each such adjustment we need to interpolate
all relevant fields onto the new grid.
In addition, the adjustment has to be carried out
several times after each individual elliptic solve, since 
we were only able to achieve a stable iteration scheme if
we pick the free constants $C_1$, $C_2$, $\omega$ and $x_{CM}$
as follows. Let us call the intersection points of the $x$-axis
with the side of the star surface not facing the origin $x_{out1}$ 
and $x_{out2}$ (located at $(A,B)=(0,0)$) for each star.
We then determine $\omega$ and $x_{CM}$ by requiring that $x_{out1}$
and $x_{out2}$ remain constant.
This task is accomplished by a root finder. In each iteration
of this root finder $C_1$, $C_2$ are adjusted such that the rest mass
of each star remains constant (by another root finder).
These root finders have to evaluate $q$ and thus need to adjust the domain
shapes several times.
If we do not adjust $\omega$ and $x_{CM}$ we find that the stars drift 
around too much for the iterations to converge.
Instead of $x_{out1}$ and $x_{out2}$ it is also possible to fix the
points $x_{max1}$ and $x_{max2}$ where the maximum values of $q$
occur.

In addition, we have observed that once a new $q$ is set
the elliptic solve often "overcorrects" which again can result
in an unstable iteration scheme. To overcome this problem
we typically do not take the $\psi$, $B^i$ and $\alpha\psi$
coming from solving Eq.~(\ref{ham_mom_dtK0}) as our new fields.
Rather, we take the average of this solution and 
the $\psi$, $B^i$ and $\alpha\psi$ from the previous iteration step
as our new fields. In this way $\psi$, $B^i$ and $\alpha\psi$
change less from one iteration step to the next.

\section{Results}
\label{results}

All the results presented in this section were computed for $n=1$
polytopes (see Eq.~(\ref{polytrop})) in units where $\kappa=1$.

In order to check that our SGRID code is working properly 
we have checked the convergence of the constraints.
For these tests we have computed the constraints directly
from Eq.~(\ref{ham0}) for different numbers of grid points.
\begin{figure}
\includegraphics[scale=0.35,clip=true]{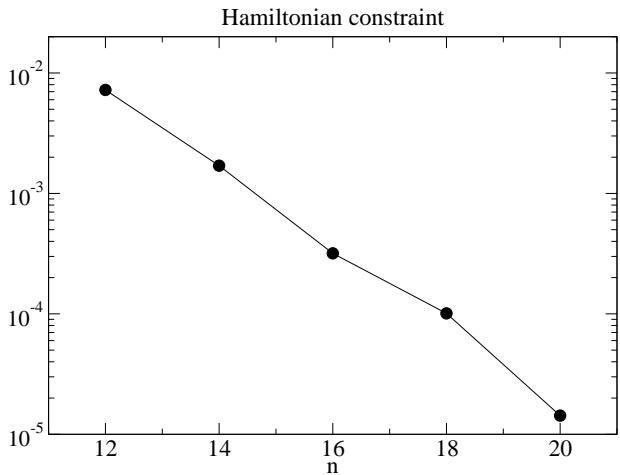}
\caption{\label{ham_conv}
The $L^2$-norm of the Hamiltonian constraint on the grid inside star 1
converges exponentially with the number of grid points $n_A=n_B=n$.
The plot is for an equal mass binary with rest masses
$m_{01}=m_{02}=0.05952$. The star centers are located 
at $x_{max1}=-x_{max2}=2.112$, and $n_{\phi}=8$ and $n_c=6$ are kept fixed.
}
\end{figure}
From Fig.~\ref{ham_conv} we see that the Hamiltonian constraint converges
exponentially with the number of grid points, as expected for
a spectral method. The momentum constraints as well as $\partial_t K$
converge to zero in a similar fashion.

\begin{table}
{\small
\begin{tabular}{|l|l|l|l|l|}
\hline
$m_{01}$	& 0.05952	& 0.1400	& 0.140		& 0.150\\
$m_{02}$	& 0.05952	& 0.0600	& 0.100		& 0.050\\
$b$		& 1.8412	& 1.8400	& 10.00		& 5.00\\
\hline
$M_{ADM}$	& 0.11572	& 0.18871	& 0.2263	& 0.1881\\
$J_{ADM}$ 	& 0.02315	& 0.04233	& 0.122		& 0.0527\\
$\omega$	& 0.038		& 0.048		& 0.0052	& 0.013\\
$d_{12}$	& 4.224		& 4.174		& 20.09		& 10.20\\
$d_1$		& 2.276		& 1.711		& 1.70		& 1.63\\
$d_2$		& 2.276		& 2.398		& 1.96		& 2.25\\
$x_{CM}$	& 0		& 0.74		& 1.6		& 2.4\\
$q_{max1}$	& 0.0285	& 0.106		& 0.108		& 0.127\\
$q_{max2}$	& 0.0285	& 0.0282	& 0.0588	& 0.0235\\
$x_{in1}$	&$+0.975$	&$+1.174$	&$+9.19	$	&$+4.25$\\
$x_{max1}$	&$+2.112$	&$+2.029$	&$+10.04$	&$+5.07$\\
$x_{out1}$	&$+3.251$	&$+2.885$	&$+10.89$	&$+5.88$\\
$x_{in2}$	&$-0.975$	&$-0.927$	&$-9.07	$	&$-4.00$\\
$x_{max2}$	&$-2.112$	&$-2.145$	&$-10.05$	&$-5.13$\\
$x_{out2}$	&$-3.251$	&$-3.325$	&$-11.03$	&$-6.25$\\
\hline\hline
$\frac{M_{ADM}}{M_{\odot}}\left(\frac{\kappa_0}{\kappa}\right)^{n/2}$
		& 1.7524	& 2.8578	& 3.427	 	& 2.849\\
\hline
$m_{01}/M_{ADM}$& 0.5143	& 0.7419	& 0.619		& 0.797\\
$m_{02}/M_{ADM}$& 0.5143	& 0.3179	& 0.442		& 0.266\\
$d_{12}/M_{ADM}$& 36.50		& 22.12		& 88.78		& 54.23\\
$d_1/M_{ADM}$	& 19.67		& 9.067		& 7.51		& 8.67\\
$d_2/M_{ADM}$	& 19.67		& 12.71		& 8.66		& 12.0\\
$\omega M_{ADM}$& 0.0044	& 0.0091	& 0.0012	& 0.0024\\
$J_{ADM}/M_{ADM}^2$&1.729	& 1.189		& 2.382		& 1.489\\
\hline
\end{tabular}
}
\caption{\label{configs}
Properties of initial data for different parameters 
$m_{01}$, $m_{02}$ and $b$. The numbers are first given in
units of $G=c=\kappa=1$. The total ADM mass $M_{ADM}$ is also
given in solar masses where $\kappa_0=5\times 10^8\mbox{m}^2$ and $n=1$
(Note that $\kappa^{-n/2} G M_{ADM}/c^2$ is dimensionless).
After that we also list some quantities in geometric units ($G=c=1$)
in terms of the total ADM mass.
}
\end{table}
We have computed initial data for the configurations
listed in table~\ref{configs}. 
Each configuration is described by the rest masses $m_{01}$
and $m_{02}$ of the two stars
given by
\begin{equation}
\label{restmass}
m_{0i} = \int_{\mbox{star} i} \rho_0 u^0 \alpha \psi^6 d^3x, \ \ \ i=\{1,2\}
\end{equation}
and the separation parameter $b$, which appears
in the coordinate transformations and is approximately 
half the separation. For each configuration we have also
computed the ADM mass and angular momentum given by
\begin{equation}
\label{MADM}
M_{ADM}= \int\left(
 \rho + \frac{1}{64\pi\alpha^2} (\bar{L}B)^{ij}(\bar{L}B)_{ij}
         \right)\psi^5 d^3x
\end{equation}
and
\begin{equation}
\label{JADM}
J_{ADM}= \int \left[(x-x_{CM}) j^y - y j^x\right]\psi^{10} d^3x .
\end{equation}
Our iterative scheme also yields the orbital angular velocity $\omega$ and
the location of the center of mass $x_{CM}$. Furthermore we list 
the maximum values of $q$ in each star along the $x$-axis,
together with their $x$-coordinates, and also the locations
of the inner and outer edges of each star.
We also show the distance $d_{12}$ between the stars and
star diameters $d_{1/2}$ defined by
$d_{12} = |x_{max1}-x_{max2}|$
and $d_{1/2} = |x_{out1/2}-x_{in1/2}|$.
The equal mass configuration in table~\ref{configs} is very
close to configurations already computed by 
Baumgarte et al.~\cite{Baumgarte:1997eg}
and also by Gourgoulhon et al.~\cite{Gourgoulhon:2000nn} 
and agrees with those to better than 
1\% (see Table II. in~\cite{Gourgoulhon:2000nn}).
Note that our code has no problems handling
unequal mass systems, as well as systems that are far apart.
In addition, it is very memory efficient. A typical run
with $n_A = n_B = 18$ points needs only about 80MB
of memory. 
The initial data can thus be generated on ordinary
PCs. On a 2.3GHz Linux PC it takes about 30 hours to push the Hamiltonian
constraint down to $10^{-4}$ if we use $n_A = n_B = 18$ points.
The main reason for the low memory footprint is that our spectral
code needs only very few grid points to achieve the
quoted accuracies.

\begin{figure}
\includegraphics[scale=0.9,clip=true, bb=80 60 575 275]
{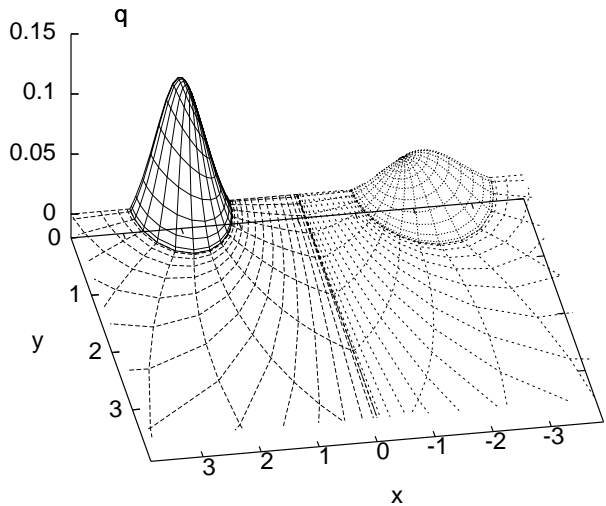}
\caption{\label{q-plot}
$q$ in the $xy$-plane for a binary with 
rest masses $m_{01}=0.14$, $m_{02}=0.06$ and $b=1.84$.
}
\end{figure}
Figure~\ref{q-plot} shows $q$ in the $xy$-plane 
for a binary with rest masses $m_{01}=0.14$ and
$m_{02}=0.06$. We can see how the domain boundaries are adapted
such that $q$ is non-zero only in the inner domains.
Note that $q$ is the rest mass density for $\kappa=n=1$.
\begin{figure}
\includegraphics[scale=0.9,clip=true, bb=80 60 575 275]
{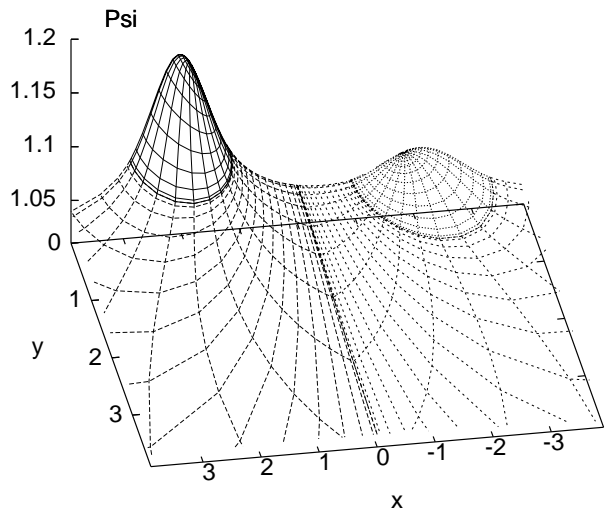}
\caption{\label{Psi-plot}
Conformal factor $\psi$ in the $xy$-plane
for a binary with rest masses $m_{01}=0.14$ and $m_{02}=0.06$.
}
\end{figure}
\begin{figure}
\includegraphics[scale=0.9,clip=true, bb=80 60 575 275]
{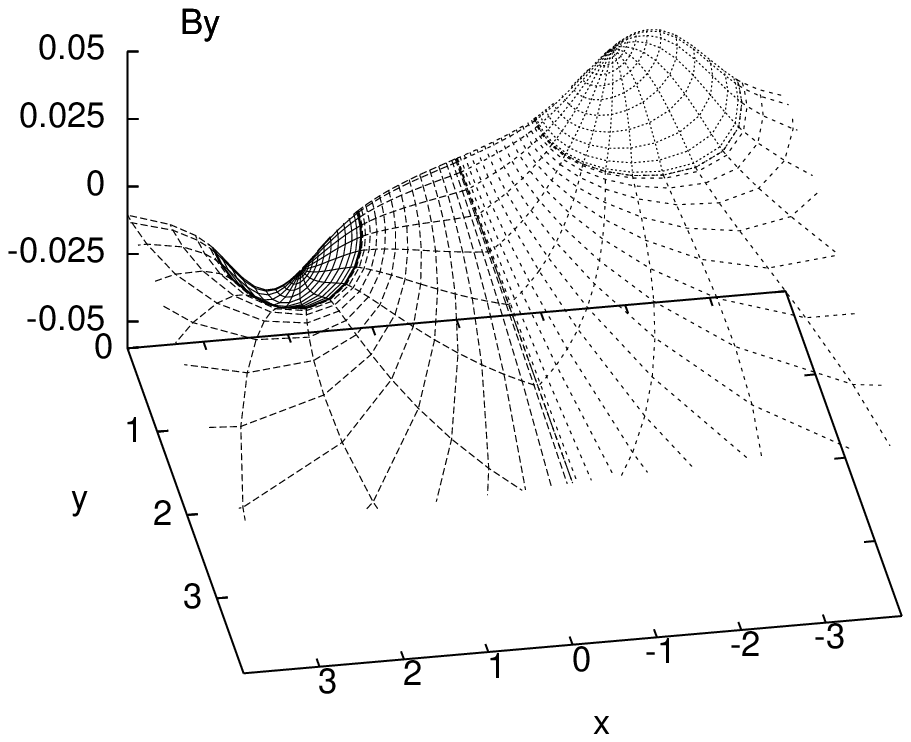}
\caption{\label{By-plot}
Largest shift component $B_y$ in inertial coordinates 
for a binary with rest masses $m_{01}=0.14$ and $m_{02}=0.06$.
}
\end{figure}
In Figs.~\ref{Psi-plot} and ~\ref{By-plot} we show 
the conformal factor $\psi$ and the largest shift component
$B_y$ for the same configuration. Note that unlike $q$ both are 
smooth ($C^{1}$) across the domain boundaries.

In order to further verify our code we have performed a comparison with
previous results from Taniguchi et al.~\cite{Taniguchi:2002ns}.
\begin{figure}
\includegraphics[scale=0.35,clip=true]{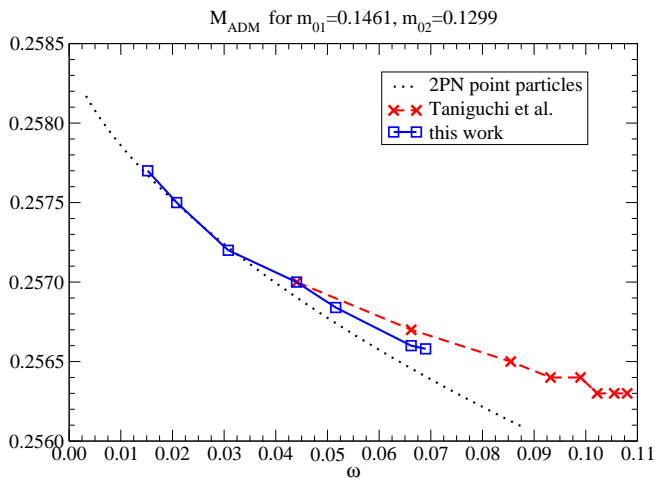}
\caption{\label{M_ADM-seq}
The ADM mass for a binary with
rest masses $m_{01}=0.1461$ and $m_{02}=0.1299$
as a function of angular velocity in units of $G=c=\kappa=1$.
Shown are results for post-2-Newtonian point particles (dotted line),
values from previous work~\cite{Taniguchi:2002ns} (crosses),
and results from our new code (squares).
}
\end{figure}
\begin{figure}
\includegraphics[scale=0.35,clip=true]{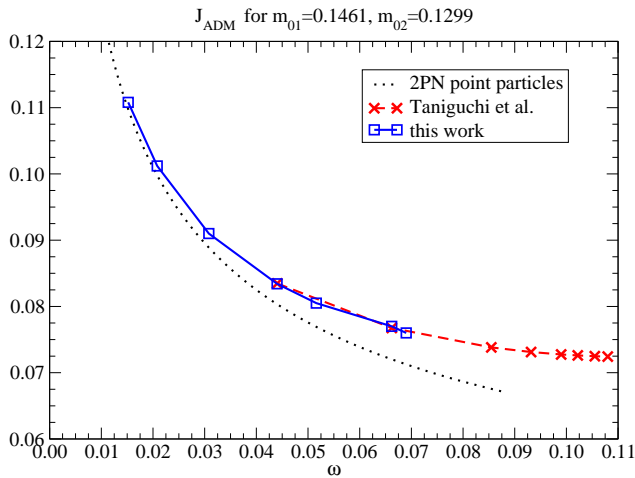}
\caption{\label{J_ADM-seq}
The ADM angular momentum for the same cases as in Fig.~\ref{M_ADM-seq}
in units of $G=c=\kappa=1$.
}
\end{figure}
In Figs.~\ref{M_ADM-seq} and ~\ref{J_ADM-seq} we show
how $M_{ADM}$ and $J_{ADM}$ vary as a function of $\omega$
for a binary with rest masses $m_{01}=0.1461$ and $m_{02}=0.1299$.
As we can see our results (squares) approach the expected
post-Newtonian results
(taken from~\cite{Schaefer93,Tichy02,Tichy03a,Tichy:2003qi}) 
for point particles (dotted line) for small
$\omega$. At intermediate $\omega$ our results
differ from point particle results and instead agree with previous 
results obtained by Taniguchi et al.~\cite{Taniguchi:2002ns}.
Note that, while the agreement in $M_{ADM}$ does not look as good
as for $J_{ADM}$, the values for $M_{ADM}$ from both methods still agree
to better than 0.05\%. 
With our current code we can construct initial data only up to
$\omega\sim0.07$. Beyond that point, already the first iteration of
our elliptic solver fails. We suspect that our initial guess
of simply using two spherical TOV stars with $B^i=0$ is not good enough
for close configurations, and that the solver would succeed if
we provided a guess that is closer to the true solution.
Notice that while Taniguchi et al.~\cite{Taniguchi:2002ns} can extend
their sequence to higher $\omega$ they also did not record
a turning point in either curve for this configuration.

\section{Discussion}
\label{discussion}

The purpose of this paper is to introduce a new numerical method
for the computation of binary neutron star initial data with
the SGRID code~\cite{Tichy:2006qn}. The method uses six domains
with different coordinate systems in each domain.
The coordinates in four of these domains
are closely related to the ones suggested in~\cite{Ansorg:2006gd}.
We have, however, added two extra domains with Cartesian coordinates
to remove coordinate singularities.
All our fields are $C^{\infty}$ inside each domain. This
allows us to use an efficient pseudo-spectral collocation method
to solve the elliptic equations (\ref{ham_mom_dtK0}) associated with
the initial data construction. Note that, we directly solve
the 5 Eqs.~(\ref{ham_mom_dtK0}), i.e. we do not split the shift
in the momentum constraint into a vector and a gradient of a scalar, which
would introduce an additional elliptic equation.
Thus we have one less equation to solve than in the original
Wilson-Mathews approach~\cite{Wilson95,Wilson:1996ty}.
Since two of our domains extend to spatial infinity we are able
to easily impose the boundary conditions in Eq.~(\ref{psi_B_alpha_BCs})
without the need of any approximations such as Robin boundary conditions.

At present we have only considered corotating
configurations. The numerical method, however, could be easily
extended to configurations with arbitrary spins, e.g. by
following the approach in~\cite{Marronetti:2003gk}.
This approach involves the addition
of one more elliptic equation for a velocity potential.
The boundary conditions for this extra equation need to
be imposed at the star surface, which is somewhat
involved if one uses cubical domains as in~\cite{Marronetti:2003gk}.
We our new method, however, such boundary conditions can
be easily imposed, since each star surface is a domain boundary.

It is well known that the star surfaces of close
configurations can develop cusps due to tidal 
forces~\cite{Uryu:1999uu,Gourgoulhon:2000nn,Marronetti:2003gk}.
We have not investigated this issue with our new method yet,
because our elliptic solver currently fails already
during the first step for close configurations.
We suspect that we need an initial guess that is
better than two spherical TOV stars with vanishing shift $B^i$.
We would like to point out, however, that our method 
should not have any additional problems with such cups if they
occur only along the $x$-axis (the line connecting the two stars).
The reason is that the $\sigma_{\pm}(B,\phi)$ which appear
in the coordinate transformations and describe the star
surfaces, can easily be chosen such that the domain
boundaries have arbitrary cusps on the $x$-axis.
Note that such cusp producing $\sigma_{\pm}(B,\phi)$ are
themselves perfectly smooth in $A,B,\phi$ coordinates, so that
we do not expect to loose spectral accuracy.

\ack
It is a pleasure to thank Pedro Marronetti
for useful discussions about NS initial data.
This work was supported by NSF grant PHY-0652874.





\vskip 1cm


\bibliographystyle{unsrt}
\bibliography{references}

\begin{thebibliography}{10}

\bibitem{LIGO:2007kva}
B.~Abbott et~al.
\newblock {LIGO: The Laser Interferometer Gravitational-Wave Observatory}.
\newblock 2007.
\newblock arXiv:0711.3041 [gr-qc].

\bibitem{LIGO_web}
http://www.ligo.caltech.edu/.

\bibitem{VIRGO_FAcernese_etal2008}
F.~Acernese et~al.
\newblock {The Virgo 3 km interferometer for gravitational wave detection}.
\newblock {\em J. Opt. A: Pure Appl. Opt.}, 10:064009, 2008.

\bibitem{VIRGO_web}
VIRGO - http://www.virgo.infn.it/.

\bibitem{GEO_web}
http://geo600.aei.mpg.de/.

\bibitem{Schutz99}
B.~Schutz.
\newblock Gravitational wave astronomy.
\newblock {\em Class. Quantum Grav.}, 16:A131--A156, 1999.

\bibitem{Peters:1963ux}
P.~C. Peters and J.~Mathews.
\newblock Gravitational radiation from point masses in a {K}eplerian orbit.
\newblock {\em Phys. Rev.}, 131:435--439, 1963.

\bibitem{Peters:1964}
P.~C. Peters.
\newblock Gravitational radiation and the motion of two point masses.
\newblock {\em Phys. Rev.}, 136:B1224--B1232, 1964.

\bibitem{Wilson95}
J.~R. Wilson and G.~J. Mathews.
\newblock Instabilities in close neutron star binaries.
\newblock {\em Phys. Rev. Lett.}, 75:4161, 1995.

\bibitem{Wilson:1996ty}
J.~R. Wilson, G.~J. Mathews, and P.~Marronetti.
\newblock {Relativistic Numerical Method for Close Neutron Star Binaries}.
\newblock {\em Phys. Rev.}, D54:1317--1331, 1996.

\bibitem{York99}
J.~W. York.
\newblock Conformal `thin-sandwich' data for the initial-value problem of
  general relativity.
\newblock {\em Phys. Rev. Lett.}, 82:1350--1353, 1999.

\bibitem{Baumgarte:1997xi}
T.~W. Baumgarte, G.~B. Cook, M.~A. Scheel, S.~L. Shapiro, and S.~A. Teukolsky.
\newblock {Binary Neutron Stars in General Relativity: Quasi- Equilibrium
  Models}.
\newblock {\em Phys. Rev. Lett.}, 79:1182--1185, 1997.

\bibitem{Baumgarte:1997eg}
T.~W. Baumgarte, G.~B. Cook, M.~A. Scheel, S.~L. Shapiro, and S.~A. Teukolsky.
\newblock {General Relativistic Models of Binary Neutron Stars in
  Quasiequilibrium}.
\newblock {\em Phys. Rev.}, D57:7299--7311, 1998.

\bibitem{Mathews:1997pf}
G.~J. Mathews, P.~Marronetti, and J.~R. Wilson.
\newblock {Relativistic Hydrodynamics in Close Binary Systems: Analysis of
  Neutron-Star Collapse}.
\newblock {\em Phys. Rev.}, D58:043003, 1998.

\bibitem{Marronetti:1998xv}
P.~Marronetti, G.~J. Mathews, and J.~R. Wilson.
\newblock {Binary neutron star systems: From the Newtonian regime to the last
  stable orbit}.
\newblock {\em Phys. Rev.}, D58:107503, 1998.

\bibitem{Bonazzola:1998yq}
Silvano Bonazzola, Eric Gourgoulhon, and Jean-Alain Marck.
\newblock {Numerical models of irrotational binary neutron stars in general
  relativity}.
\newblock {\em Phys. Rev. Lett.}, 82:892--895, 1999.

\bibitem{Gourgoulhon:2000nn}
Eric Gourgoulhon, Philippe Grandclement, Keisuke Taniguchi, Jean-Alain Marck,
  and Silvano Bonazzola.
\newblock {Quasiequilibrium sequences of synchronized and irrotational binary
  neutron stars in general relativity. I. Method and tests}.
\newblock {\em Phys. Rev.}, D63:064029, 2001.

\bibitem{Marronetti:1999ya}
P.~Marronetti, G.~J. Mathews, and J.~R. Wilson.
\newblock Irrotational binary neutron stars in quasi-equilibrium.
\newblock {\em Phys. Rev. D}, 60:087301, 1999.

\bibitem{Uryu:1999uu}
Koji Uryu and Yoshiharu Eriguchi.
\newblock {A new numerical method for constructing quasi-equilibrium sequences
  of irrotational binary neutron stars in general relativity}.
\newblock {\em Phys. Rev.}, D61:124023, 2000.

\bibitem{Marronetti:2003gk}
Pedro Marronetti and Stuart~L. Shapiro.
\newblock {Relativistic models for binary neutron stars with arbitrary spins}.
\newblock {\em Phys. Rev.}, D68:104024, 2003.

\bibitem{Taniguchi:2002ns}
Keisuke Taniguchi and Eric Gourgoulhon.
\newblock {Quasiequilibrium sequences of synchronized and irrotational binary
  neutron stars in general relativity. III: Identical and different mass stars
  with gamma = 2}.
\newblock {\em Phys. Rev.}, D66:104019, 2002.

\bibitem{Taniguchi:2003hx}
Keisuke Taniguchi and Eric Gourgoulhon.
\newblock {Various features of quasiequilibrium sequences of binary neutron
  stars in general relativity}.
\newblock {\em Phys. Rev.}, D68:124025, 2003.

\bibitem{Tichy:2006qn}
Wolfgang Tichy.
\newblock Black hole evolution with the bssn system by pseudo-spectral methods.
\newblock {\em Phys. Rev.}, D74:084005, 2006.

\bibitem{Misner73}
C.~W. Misner, K.~S. Thorne, and J.~A. Wheeler.
\newblock {\em Gravitation}.
\newblock W. H. Freeman, San Francisco, 1973.

\bibitem{Tichy02}
Wolfgang Tichy, Bernd Br\"ugmann, Manuela Campanelli, and Peter Diener.
\newblock Binary black hole initial data for numerical general relativity based
  on post-{N}ewtonian data.
\newblock {\em Phys. Rev. D}, 67:064008, 2003.
\newblock gr-qc/0207011.

\bibitem{Kelly:2007uc}
Bernard~J. Kelly, Wolfgang Tichy, Manuela Campanelli, and Bernard~F. Whiting.
\newblock Black hole puncture initial data with realistic gravitational wave
  content.
\newblock {\em Phys. Rev.}, D76:024008, 2007.

\bibitem{Lightman75}
Alan~P. Lightman, William~H. Press, Richard~H. Price, and Saul~A. {T}eukolsky.
\newblock {\em Problem Book in Relativity and Gravitation}.
\newblock Princeton University Press, Princeton, NJ, 1975.

\bibitem{Ansorg:2004ds}
Marcus Ansorg, Bernd Br\"ugmann, and Wolfgang Tichy.
\newblock A single-domain spectral method for black hole puncture data.
\newblock {\em Phys. Rev. D}, 70:064011, 2004.

\bibitem{Ansorg:2005bp}
Marcus Ansorg.
\newblock {A double-domain spectral method for black hole excision data}.
\newblock {\em Phys. Rev.}, D72:024018, 2005.

\bibitem{Ansorg:2006gd}
Marcus Ansorg.
\newblock {Multi-Domain Spectral Method for Initial Data of Arbitrary Binaries
  in General Relativity}.
\newblock {\em Class. Quant. Grav.}, 24:S1--S14, 2007.

\bibitem{Bonazzola:1998qx}
Silvano Bonazzola, Eric Gourgoulhon, and Jean-Alain Marck.
\newblock {Numerical approach for high precision 3-D relativistic star models}.
\newblock {\em Phys. Rev.}, D58:104020, 1998.

\bibitem{Davis-Duff-1997-UMFPACK}
Timothy~A. Davis and Iain~S. Duff.
\newblock An unsymmetric-pattern multifrontal method for sparse {LU}
  factorization.
\newblock {\em SIAM J. Matrix Anal. Applic.}, 18(1):140--158, 1997.

\bibitem{Davis-Duff_UMFPACK_1999}
Timothy~A. Davis and Iain~S. Duff.
\newblock A combined unifrontal/multifrontal method for unsymmetric sparse
  matrices.
\newblock {\em ACM Trans. Math. Softw.}, 25(1):1--20, 1999.

\bibitem{Davis_UMFPACK_V4.3_2004}
Timothy~A. Davis.
\newblock Algorithm 832: Umfpack v4.3---an unsymmetric-pattern multifrontal
  method.
\newblock {\em ACM Trans. Math. Softw.}, 30(2):196--199, 2004.

\bibitem{Davis_UMFPACK_2004}
Timothy~A. Davis.
\newblock A column pre-ordering strategy for the unsymmetric-pattern
  multifrontal method.
\newblock {\em ACM Trans. Math. Softw.}, 30(2):165--195, 2004.

\bibitem{umfpack_web}
Timothy~A. Davis.
\newblock {UMFPACK} a sparse linear systems solver using the Unsymmetric
  MultiFrontal method:\\ {\tt
  http://www.cise.ufl.edu/research/sparse/umfpack/}.

\bibitem{Schaefer93}
G.~Sch\"afer and N.~Wex.
\newblock {\em Physics Lett. A}, 174:196, 1993.

\bibitem{Tichy03a}
Wolfgang Tichy, Bernd Br\"ugmann, and Pablo Laguna.
\newblock Gauge conditions for binary black hole puncture data based on an
  approximate helical {K}illing vector.
\newblock {\em Phys. Rev. D}, 68:064008, 2003.

\bibitem{Tichy:2003qi}
Wolfgang Tichy and Bernd Br{\"u}gmann.
\newblock Quasi-equilibrium binary black hole sequences for puncture data
  derived from helical killing vector conditions.
\newblock {\em Phys. Rev. D}, 69:024006, 2004.

\end{thebibliography}

\end{document}